\documentclass[12pt]{iopart}

\usepackage{epsfig,amssymb} 

 \def\be{\begin{equation}}
\def\ee{\end{equation}}    
\def\baray{\begin{eqnarray}}
\def\earay{\end{eqnarray}}

\def\2pi{\left(2\pi\right)}

\def\hf{\frac{1}{2}}

\def\pq{\left(p,q\right)}

\begin{document}
\title{Dynamics of F/D networks: the r\^ole of bound states}
\author{Mairi Sakellariadou$^1$ and Horace Stoica$^2$}
\address{$^1$ Department of Physics,  
King's College London, 
London WC2R 2LS, U.K.} 
\ead{Mairi.Sakellariadou@kcl.ac.uk}
\address{$^2$ Blackett Laboratory, 
Imperial College London, 
SW7 2AZ U.K.}
\eads{\mailto{mairi.sakellariadou@kcl.ac.uk},\;\mailto{f.stoica@imperial.ac.uk}}

\date{\today}

\begin{abstract}

We study, via numerical experiments, the r\^ole of bound states in the
evolution of cosmic superstring networks, being composed by $p$
F-strings, $q$ D-strings and $(p,q)$ bound states. We find robust
evidence for scaling of all three components of the network,
independently of initial conditions.  The novelty of our numerical
approach consists of having control over the initial abundance of
bound states. This indeed allows us to identify the effect of bound
states on the evolution of the network. Our studies also clearly show
the existence of an additional energy loss mechanism, resulting to a
lower overall string network energy, and thus scaling of the
network. This new mechanism consists of the formation of bound states
with an increasing length.
\end{abstract}
\submitto{JCAP}



\section{Introduction}
 
Cosmic superstrings can be formed as a result of brane annihilations
in the context of brane-world
scenario~\cite{Sarangi:2002yt,Jones:2003da,dvalietal}. In the brane-world
models, brane inflation takes place~\cite{dvali-tye99} while two
branes move towards each other, and their annihilation releases the
brane tension energy that heats up the universe to start the hot big
bang era. Typically, strings of all sizes and types may be produced
during the collision. Considering a IIB string theory in a
(9+1)-dimensional space-time, interactions of Dirichlet (D) branes
leads to the {\sl unwinding} and subsequent {\sl evaporation} of
higher dimensionality branes, with the survival of three-dimensional
(D3) branes embedded in a (9+1)-dimensional bulk --- one of which
could play the r\^ole of our universe --- and D1-branes
(D-strings)~\cite{Durrer:2005nz}. Large Fundamental (F) strings and
D-strings that survive the cosmological evolution become cosmic
superstrings.  They are of cosmological size and could play the r\^ole
of cosmic strings~\cite{Vilenkin_shellard,ms-cs07}, false vacuum
remnants formed generically at the end of hybrid inflation within
Grand Unified Theories~\cite{jrs03,ms-cs08}. Cosmic superstrings have
gained a lot of interest, particularly since it is believed that they
may be observed in the sky, providing both, a means of testing string
theory and a hint for a physically motivated inflationary model (for a
recent review, see e.g. Ref.~\cite{sakellariadou2008}).

Brane collisions lead also to the formation of bound states,
$(p,q)$-strings, which are composites of $p$ F-strings and $q$
D-strings~\cite{cmp04,lt04}.  The presence of stable bound states
implies the existence of junctions, where two different types of
string meet at a point and form a bound state leading away from that
point.  Thus, when cosmic superstrings of different types collide,
they can not intercommute, instead they exchange partners and form a
$Y$-junction, as a consequence of charge conservation at the junction
of colliding $(p, q)$-strings.  The evolution of F, D strings and
their bound states is a rather complicated problem, which necessitates
both numerical as well as analytical investigations. Junctions may
prevent the network from achieving a {\sl scaling} solution,
invalidating the cosmological model leading to their formation.  In a
number of studies, cosmic superstring evolution has been addressed via
numerical
experiments~\cite{Sakellariadou:2004wq,Avgoustidis:2004zt,Copeland:2005cy,Hindmarsh:2006qn,Rajantie:2007hp,Urrestilla:2007yw}.
The formation of three-string junctions and kinematic constraints for
their collisions have been also investigated
analytically~\cite{cks06,Copeland:2006if,Copeland:2007nv}.

In what follows, we address the question of the effect of junctions in
the evolution of a cosmic superstring network, being composed by three
components: $p$ F-strings, $q$ D-strings and their $(p,q)$ bound
states. The results of our numerical investigations can be summarised
as follows. Firstly, there is clear evidence for {\sl scaling} of all
three components of the network, independently of the chosen initial
configurations. Secondly, the existence of bound states effects the
evolution of the network. Thirdly, for $(p,q)$ strings there is a
supplementary energy loss mechanism, in addition to the chopping off
of loops, and it is this new mechanism that allows the network to
scale. More precisely, the additional energy loss mechanism is the
formation of bound states, whose length increases, lowering the
overall energy of the network. We note that in our simulations we can
have control over the initial population of bound states, and this
renders our novel results particularly important.

\section{The model}
We adopt the model of Ref.~\cite{Rajantie:2007hp}. This is a simple
field theory model of $\pq$ bound states, in analogy to the Abelian
Higgs model, which incorporates the main features of string theory.
To represent the two different species of strings, the model includes
two complex scalar fields, $\phi$ and $\chi$, of the Abelian Higgs
model, coupled via a potential, so that bound states can be formed.
The presence of stable bound states implies the existence of
junctions, where two different types of string meet at a point and
form a bound state leading away from that point.

We distinguish two different cases of cosmic superstring networks.
The nature of cosmic superstrings depends on the particular brane
inflationary model leading to their formation. In the case that both
species of cosmic strings are BPS, the model is described by the
action~\cite{Rajantie:2007hp}:
 \begin{eqnarray} 
\label{equ:action}
 \mathcal{S} &=& \int {\rm d}^{3}x{\rm d}t \biggl[\biggr.
   -\frac{1}{4}F^{2} -\hf \left(D_\mu\phi\right)
   \left(D^{\mu}\phi\right)^{*}
   -\frac{\lambda_{1}}{4}\left(\phi\phi^{*}-\eta_{1}^{2}\right)^{2}
   \nonumber \\ && \ \ \ \ \ \ \ \ \ \ \ \ \ -\frac{1}{4}H^{2}
   -\hf \left(D_\mu\chi\right) \left(D^{\mu}\chi\right)^{*}
   -\frac{\lambda_{2}}{4}\phi\phi^{*}
   \left(\chi\chi^{*}-\eta_{2}^{2}\right)^{2} \biggl.\biggr]~,
\end{eqnarray} 
where the covariant derivative $D_\mu$ is defined by
\begin{eqnarray}
D_\mu\phi&=&\partial_\mu\phi-ie_1A_\mu\phi~,\nonumber\\
D_\mu\chi&=&\partial_\mu\chi-ie_2C_\mu\chi~.
\end{eqnarray}
For clarity in our discussion, we label the $\phi$ field as ``Higgs''
and the $\chi$ field as ``axion'', even though both fields are
Higgs-like.  The scalars are coupled to the U(1) gauge fields $A_\mu$
and $C_\mu$, with coupling constants $e_1$ and $e_2$ and field
strength tensors $F_{\mu\nu}=\partial_\mu A_\nu -\partial_\nu A_\mu$
and $H_{\mu\nu}=\partial_\mu C_\nu -\partial_\nu C_\mu$, respectively.
The scalar potentials are parametrised by the positive constants
$\lambda_1, \eta_1$ and $\lambda_2, \eta_2$, respectively.

In the case that one species of string is non-BPS, we remove the
second gauge field by setting $e_2=0$. In this way, this species of
string is represented by the topological defect of a complex scalar
field with a global U(1) symmetry. Such defects are characterised by
the existence of long-range interactions \cite{Moore:2006ec} 
--- as opposed to local
strings in which all energy density is confined within the string, so
that local strings have only gravitational interactions --- implying
different consequences for the evolution of the network.

In flat space the classical equations of motion for the $\phi$ and 
$\chi$ fields, obtained from the above action, Eq.~(\ref{equ:action}),
read~\cite{Rajantie:2007hp}
\begin{eqnarray}
\label{equ:eom}
\partial_\mu F^{\mu\nu}&=&2e_1{\rm Im}\phi^*
D^\nu\phi,
\nonumber\\ 
\partial_\mu H^{\mu\nu}&=&2e_2{\rm Im}\chi^*
D^\nu\chi,
\nonumber\\ 
D_{\mu} D^\mu \phi &=& -2\lambda_1
\left(\phi^*\phi -\eta_1^2\right)\phi - 
\lambda_{2}\left(\chi^{*}\chi-\eta_{2}^{2}\right)^2\phi
\nonumber\\ 
D_{\mu} D^\mu \chi
&=& -2\lambda_2 \phi^*\phi\left(\chi^*\chi -\eta_2^2\right)\chi.
\end{eqnarray}
By calculating numerically the tension of the $\pq$ strings, for a
range of values of the charges, and then fitting the data to the
square-root expression
\begin{equation} 
\label{pq_tension} 
\mu_{(p,q)}=\mu_{\rm F}\sqrt{p^2+q^2/g_{\rm s}^2}~, 
\end{equation} 
where $\mu_{\rm F}$ denotes the effective fundamental string tension
after compactification and $g_{\rm s}$ stands for the string coupling,
we check~\cite{Rajantie:2007hp} the validity of our model. We remind
to the reader that Eq.~(\ref{pq_tension}) is exact only in the BPS
limit.
 
The potential is chosen such that bound states are energetically preferred
over single strings. This can be seen from the form of the potential for 
the axion strings. At the core of the Higgs string the field vanishes, 
$\phi = 0$, and therefore the potential of the axion vanishes as well. 
Therefore it is energetically favourable for the axion string to be 
located at the core of the Higgs string. In the true vacuum of the Higgs,
we have
$\left|\phi\right| = \eta_{1}$, so the axion string behaves just like 
the Higgs string. 

We employ this model to study, via numerical simulations, overall
properties of the cosmic superstring network. In particular, we are
interested in investigating the effect of junctions and their r\^ole
in achieving {\sl scaling} of the network.
 
In the actual simulations we include the expansion of the universe by
considering strings of fixed co-moving thickness, according to the
model of Refs.~\cite{Ryden:1989vj,Press:1989yh,Moore:2001px}.  In a
Friedmann-Lema\^{i}tre-Robertson-Walker universe and considering the
temporal gauge, $A_{0}=0$, the equations of motion for the scalar
fields take the form:
\baray
\partial_{0}F_{0i} + 2\left(1-s\right)\frac{a^{\prime}}{a}F_{0i}
&=& 2a^{2s}e_1{\rm Im}\phi^*D^\nu\phi + 
\partial_{j}F_{ji}
\\ 
\partial_{0}C_{0i} + 2\left(1-s\right)\frac{a^{\prime}}{a}C_{0i}
&=& 2a^{2s}e_2{\rm Im}\chi^*D^\nu\chi + 
\partial_{j}C_{ji}
\\
\ \ \ \
\phi^{\prime\prime} + 
2\frac{a^{\prime}}{a}\phi^{\prime} + 
D_{i}D_{i}\phi &=& -a^{2s} \left[2\lambda_{1} 
\left(\phi^{*}\phi-\eta_{1}^{2}\right)
+ \lambda_{2} 
\left(\chi^{*}\chi-\eta_{2}^{2}\right)^{2}\right]\phi
\\
\ \ \ \
\chi^{\prime\prime} + 
2\frac{a^{\prime}}{a}\chi^{\prime} + 
D_{i}D_{i}\chi &=& 
-2\lambda_{2} a^{2s} \phi^{*}\phi\left(\chi^{*}\chi-\eta_{2}^{2}\right)\chi
\earay
where the time variable is the conformal time $\tau$:
\be
{\rm d}s^{2}=a^{2}\left(\tau\right)\left(-{\rm d}\tau^{2}+{\rm d}{\bf x}^{2}
\right)~.
\ee
In our simulations we set $s=0$, instead of the $s=1$ required by
the continuum-limit of the equations of motion (otherwise the defects will 
{\sl fall through the lattice}); we also take, for simplicity, 
$a^{\prime}/a = 1/\tau$.
 
We calculate the correlation length using the expression:
\be
\xi = \sqrt{\frac{V}{L}}~,
\ee
where $V$ is the volume of the simulation box and $L$ is the total
length of the string network inside the box (following the same
approach as for example in Ref.~\cite{Urrestilla:2007yw}).  We identify the
location of the string by the value of the scalar fields.

\section{Results}

We present our main results for the {\sl scaling} of $\pq$ string
networks, using simulations of the classical time evolution on
lattices with three spatial dimensions. We consider two types of
networks, namely local-global and local-local networks.  The axion
field has a global U(1) symmetry, and therefore the axion strings
display long-range interactions.

The lattice discretisation of the equations of motion is discussed in
Ref.~\cite{Rajantie:2007hp}. We choose all parameters of the Lagrangian to
have {\sl natural} values,
\begin{equation} 
\eta_1 = 1,\; \lambda_1 = 1,\; e_1 = 1,\; \eta_2 = 1,\; \lambda_2
= 1,\; e_2 = 1~.  
\end{equation} 
The lattice spacing is $\delta x=0.45$, roughly half the
characteristic length scale set by the masses of the scalar and gauge
fields in the broken phase. The time-step is $\delta t=10^{-2}$.  The
simulations are carried out with boxes of volume $256^3$.

For each of the two (local-global and local-local) networks we
consider two types of initial conditions: one in which a large
percentage of strings ($\sim$ 50 \% of the string length) is in bound
states, and the other one with a small percentage of strings ($\sim$ 2
\% of the string length) is in bound states.  We are able to achieve
the small bound state percentage configuration, by starting with a
large percentage configuration and then {\sl shifting} one of the
fields by roughly half the initial correlation length. 

The type of behaviour we observe is that in most cases the string
network does have a {\sl scaling} behaviour, but the slope, $\gamma$, 
of the correlation length, $\xi$, versus time, defined by:
\be
\xi\left(\tau\right) = \gamma\tau~,
\ee
undergoes a discrete change as the network evolves. This change
is most clearly visible in the case of a local-global string network,
for both large and small amounts of bound states. The most dramatic
change is that for the correlation length of the bound states
themselves, provided the amount of bound states is small.

For local-local networks, the change in correlation length is not
detectable if there is a large amount of bound states, while if there
is a small amount of bound states we find a case where the bound
states themselves do not exhibit {\sl scaling} behaviour.  This
violation of {\sl scaling} by the bound states can be explained by
taking into account that the formation of bound states gives an
additional possibility for a network of $\pq$ strings to decrease its
energy. Therefore, the length of the bound states increases as a
function of time, violating the {\sl scaling} behaviour which requires
the total length to decrease. Once the length of the bound states have
reach the maximum value, {\sl scaling} resumes. This violation of
{\sl scaling} by the bound states does not lead to violation of {\sl
  scaling} for the entire network, as it can be observed from the
evolution of the Higgs and axion strings.

\begin{figure}[htbp] 
  \begin{center} 
    \includegraphics[width=0.32\textwidth,angle=0]{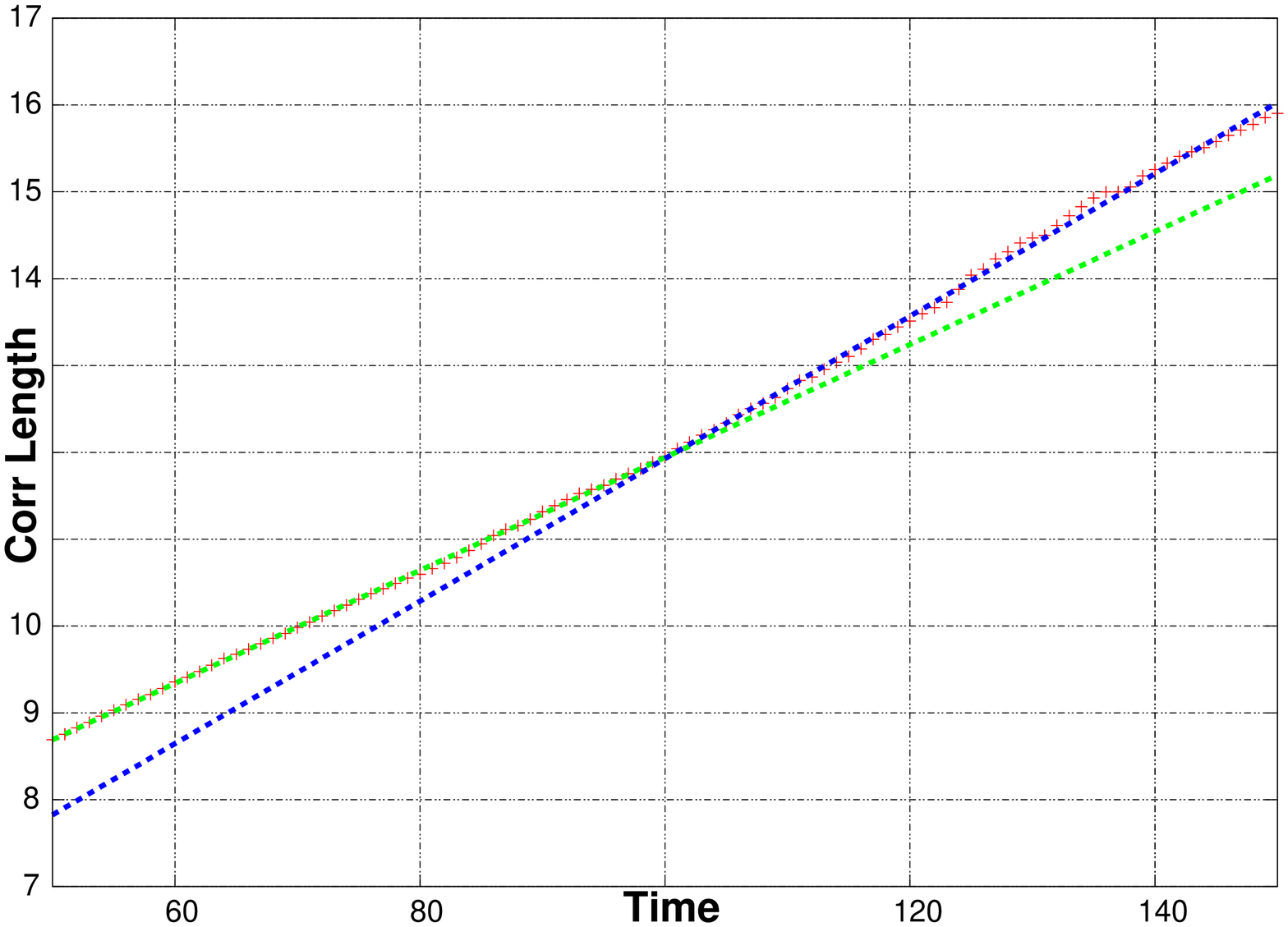}
    \includegraphics[width=0.32\textwidth,angle=0]{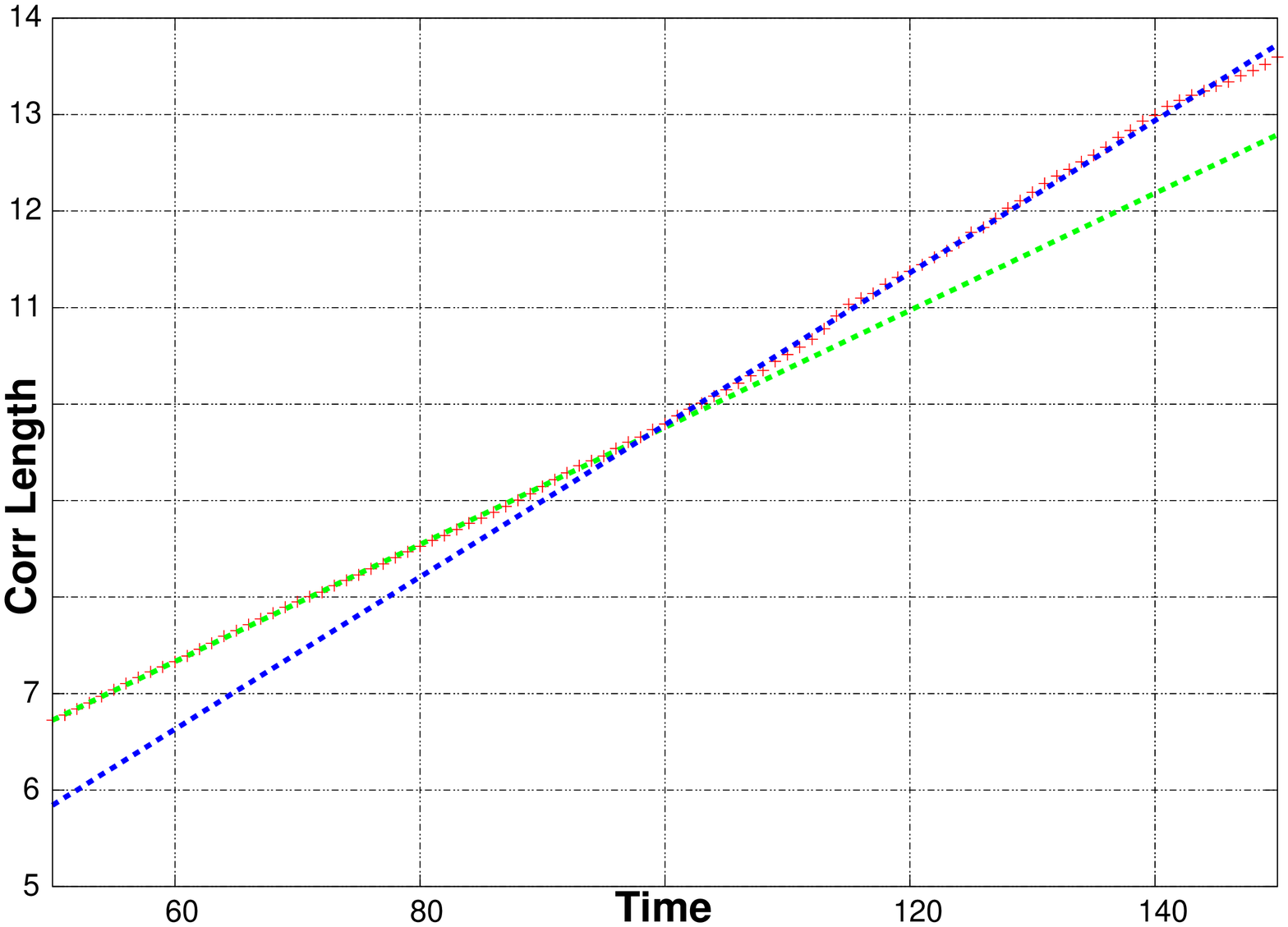}
    \includegraphics[width=0.32\textwidth,angle=0]{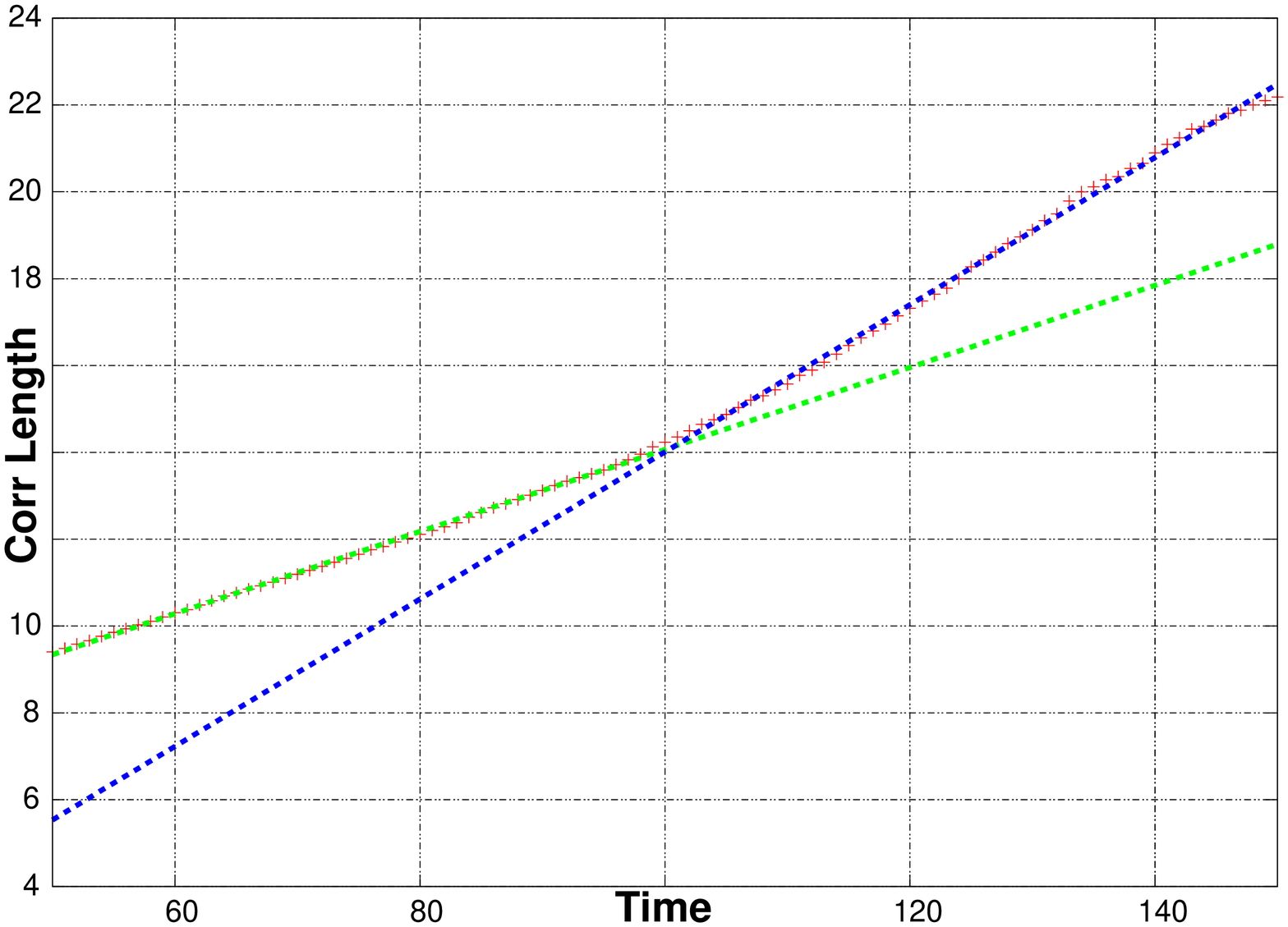}
    \caption{The Higgs (left), axion (middle), and bound state (right)
      string correlation length as a function of time. The network is
      a local-global one with a large amount of bound states. The data
      and linear fits for the two regimes are shown.
      \label{LG_High}} 
  \end{center} 
\end{figure}

In Fig.~\ref{LG_High} we show the string correlation length for the
Higgs and axion fields, as well as for their bound states, as a
function of time. The initial configuration is a local-global network
with a large amount of bound states.  Clearly, there is convincing
evidence for scaling of the three components of the network. This
scaling is characterised with a distinct change of the correlation
length slope during the network evolution.

\begin{figure}[htbp] 
  \begin{center} 
    \includegraphics[width=0.32\textwidth,angle=0]{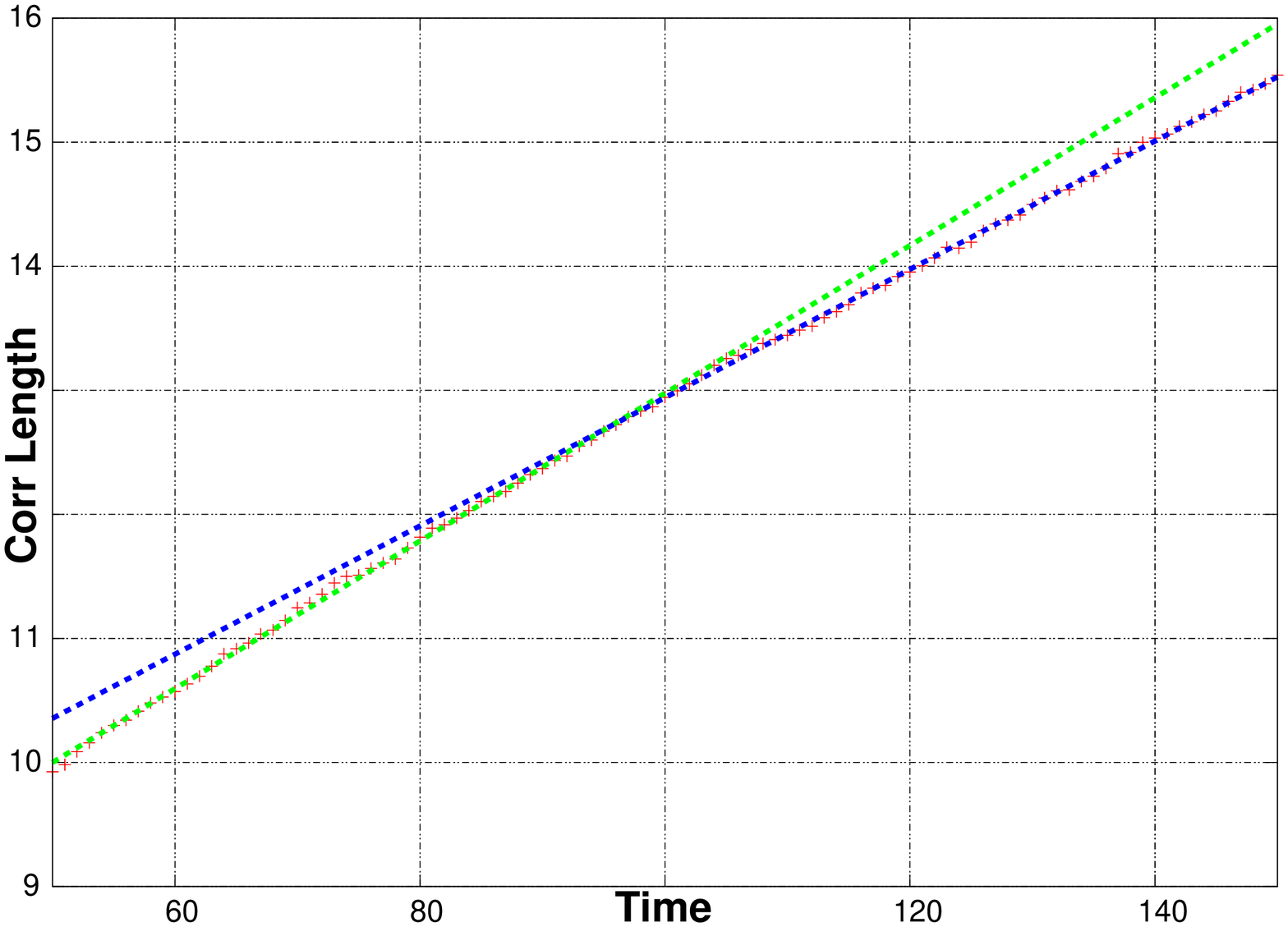}
    \includegraphics[width=0.32\textwidth,angle=0]{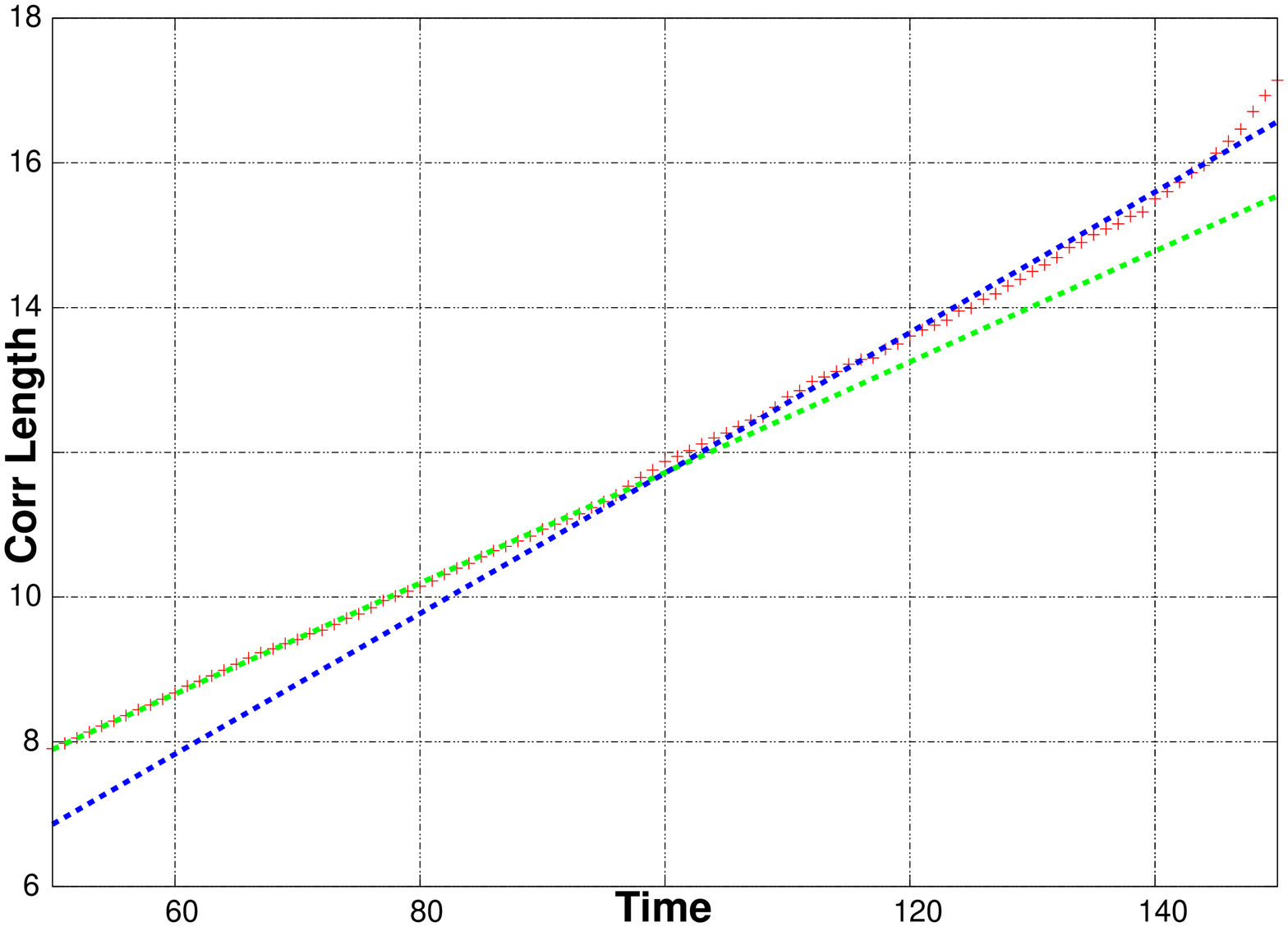}
    \includegraphics[width=0.32\textwidth,angle=0]{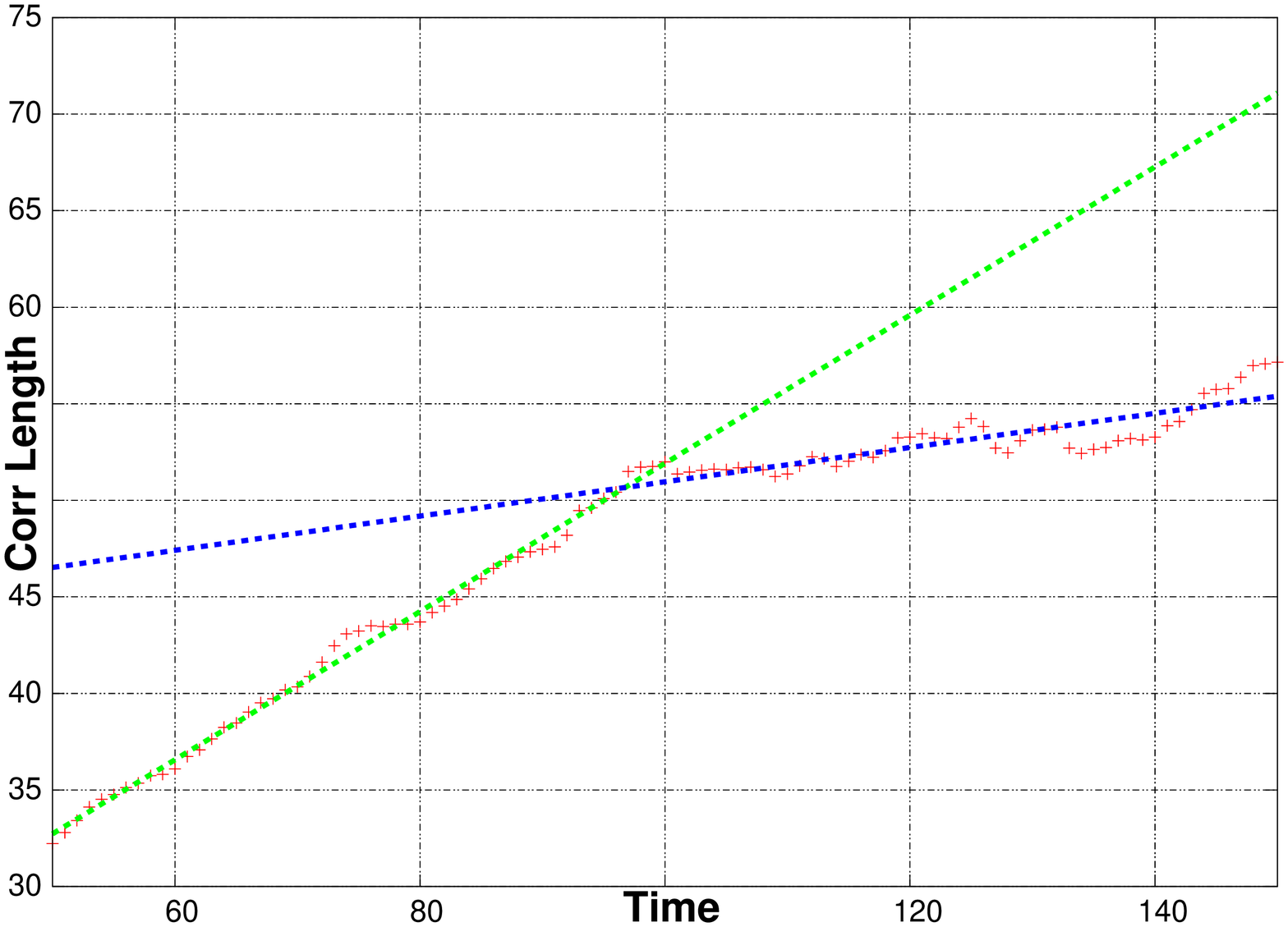}
    \caption{The Higgs (left),  axion (middle), and bound state (right) 
      string correlation length as a function of time. The network is a 
      local-global one with a small amount of bound states.  The data
      and linear fits for the two regimes are shown.
      \label{LG_Low}} 
  \end{center} 
\end{figure}

In Fig.~\ref{LG_Low} we show the string correlation length for the
Higgs and axion fields, as well as for their bound states, as a
function of time. The initial configuration is a local-global network
with a small amount of bound states. Again, scaling behaviour is
apparent for all three components of the local-global network. The
difference between the case plotted in Fig.~\ref{LG_Low} and the one
plotted in Fig.~\ref{LG_High}, is the percentage of initial bound
states. As one can easily observe from Fig.~\ref{LG_Low}, if the
initial state of a local-global network has a small amount of bound
states, then the change of the correlation length slope during the evolution
of the network is more acute.

\begin{figure}[htbp]
  \begin{center} 
    \includegraphics[width=0.32\textwidth,angle=0]{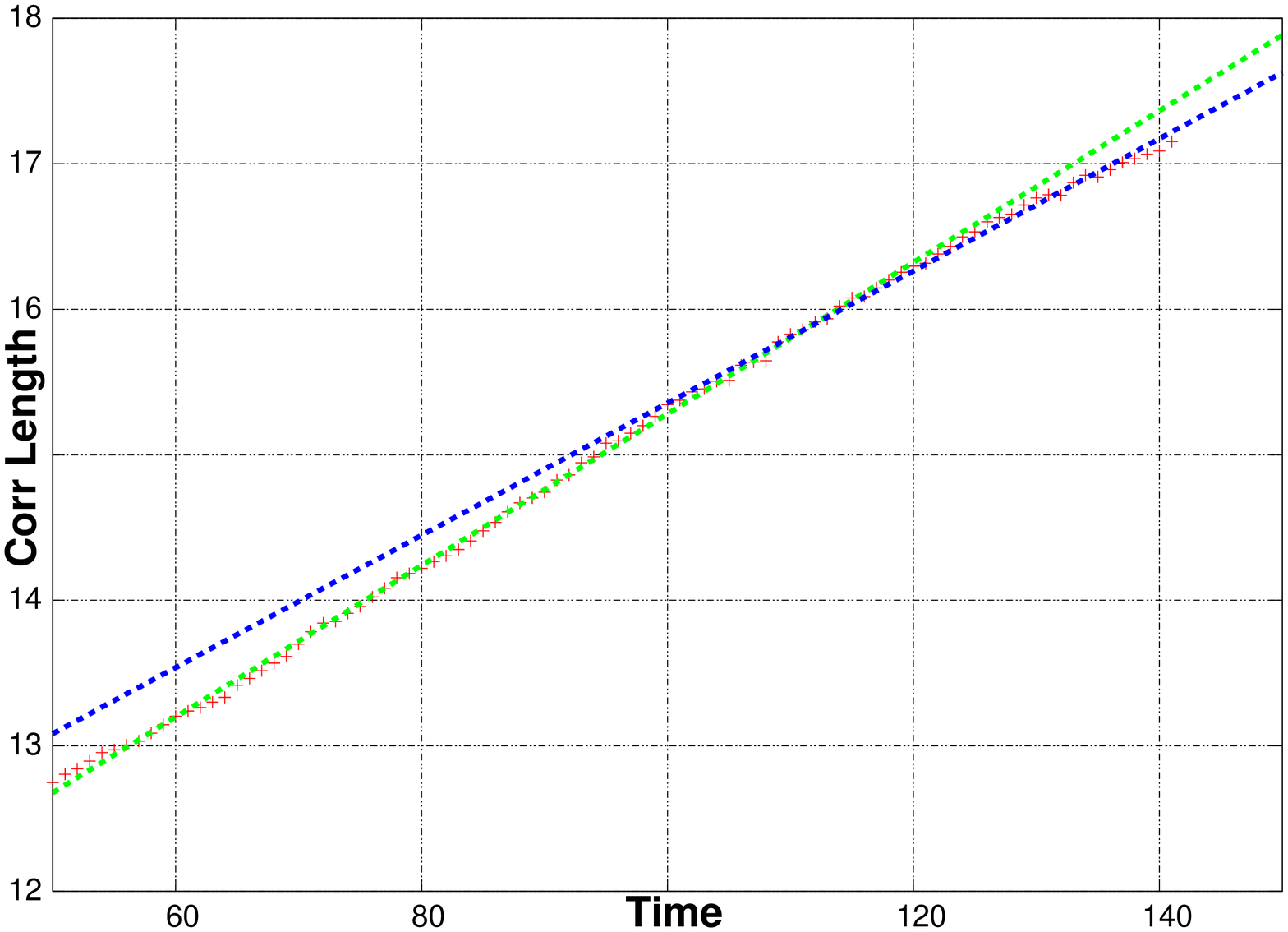}
    \includegraphics[width=0.32\textwidth,angle=0]{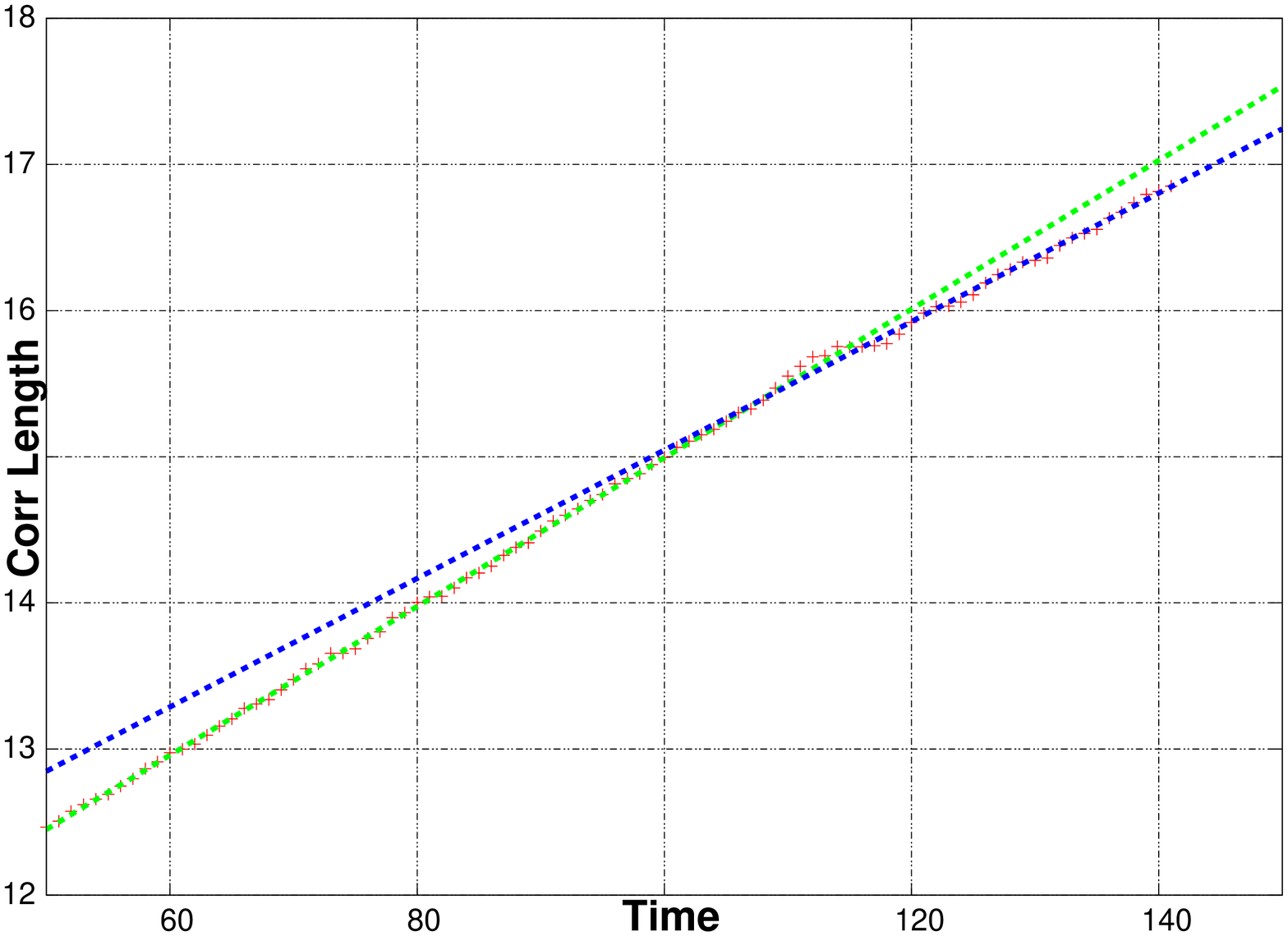}
    \includegraphics[width=0.32\textwidth,angle=0]{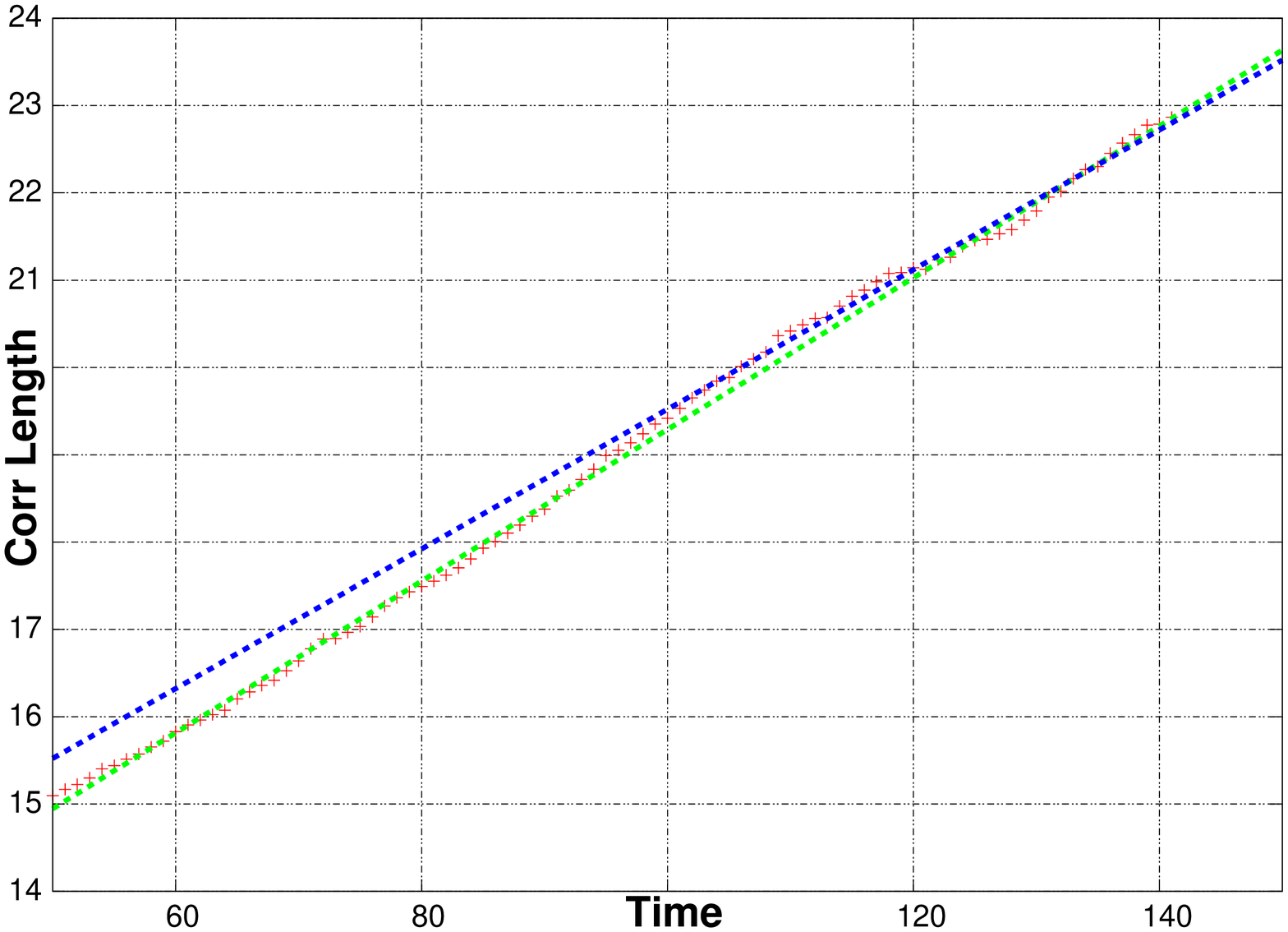}
    \caption{The Higgs (left),  axion (middle), and bound state (right) 
      string correlation length as a function of time. The network is a 
      local-local one with a large amount of bound states.  The data
      and linear fits for the two regimes are shown.
      \label{LL_High}} 
  \end{center} 
\end{figure}

\begin{figure}[htbp] 
  \begin{center} 
    \includegraphics[width=0.32\textwidth,angle=0]{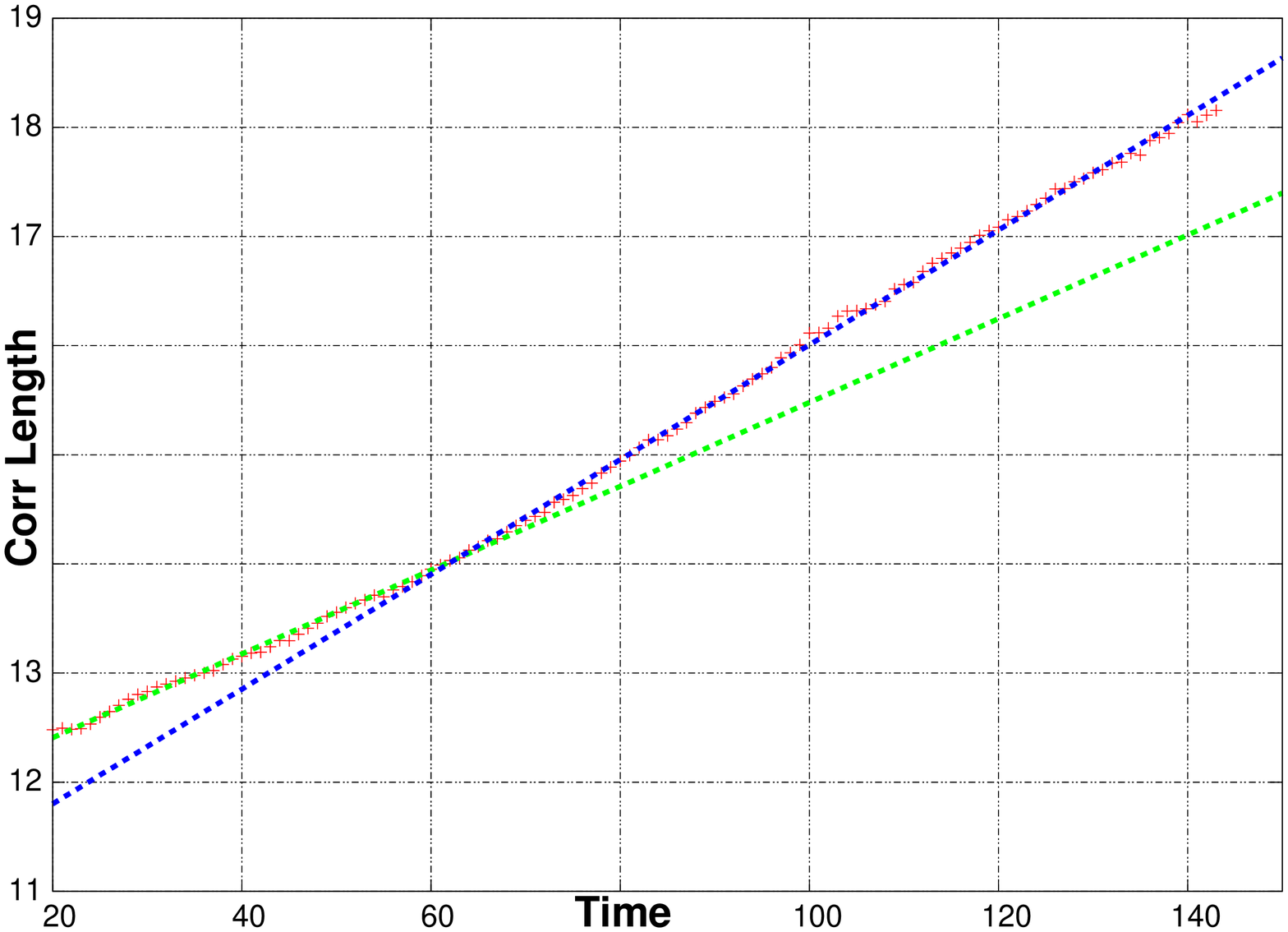}
    \includegraphics[width=0.32\textwidth,angle=0]{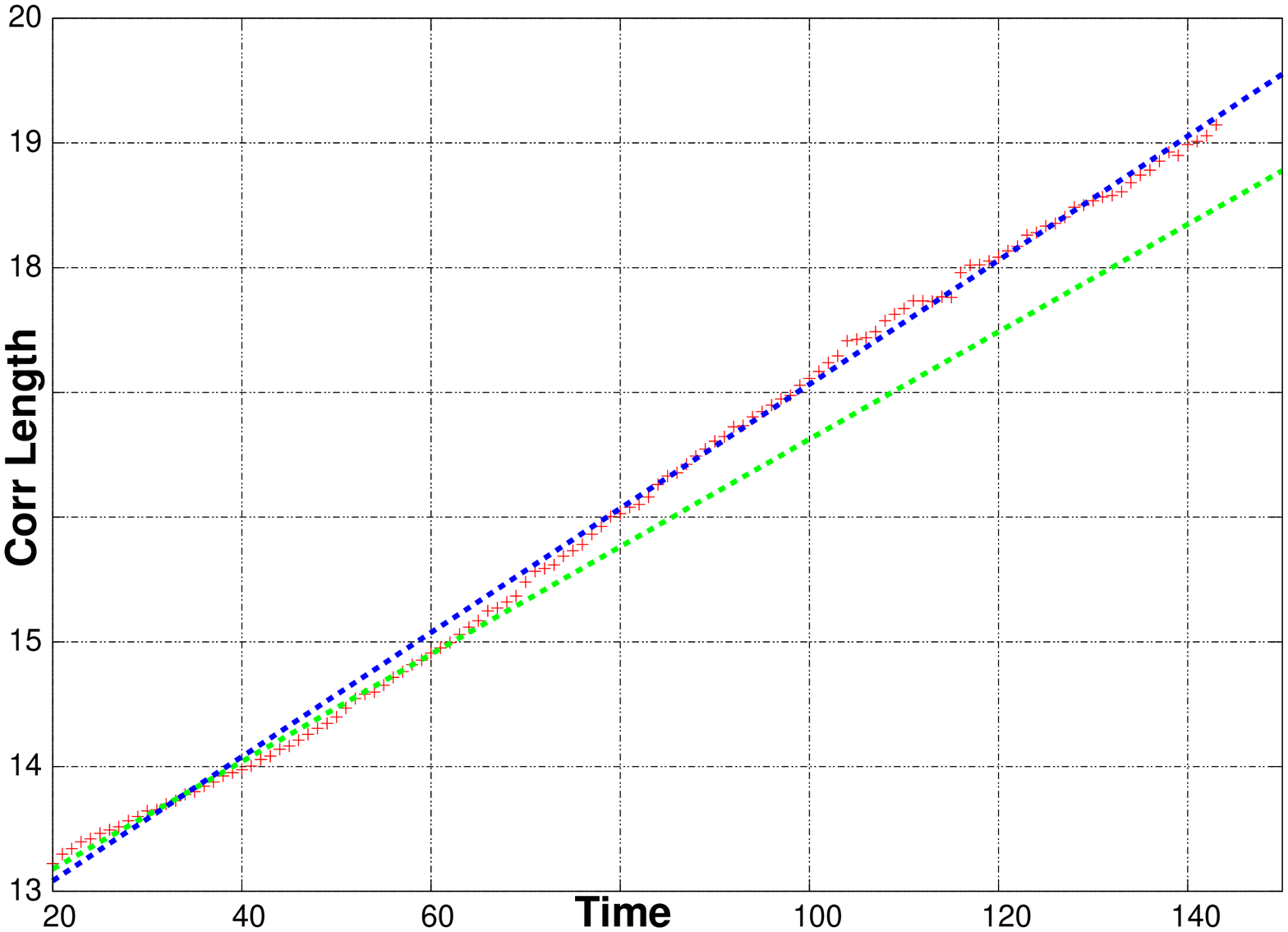}
    \includegraphics[width=0.32\textwidth,angle=0]{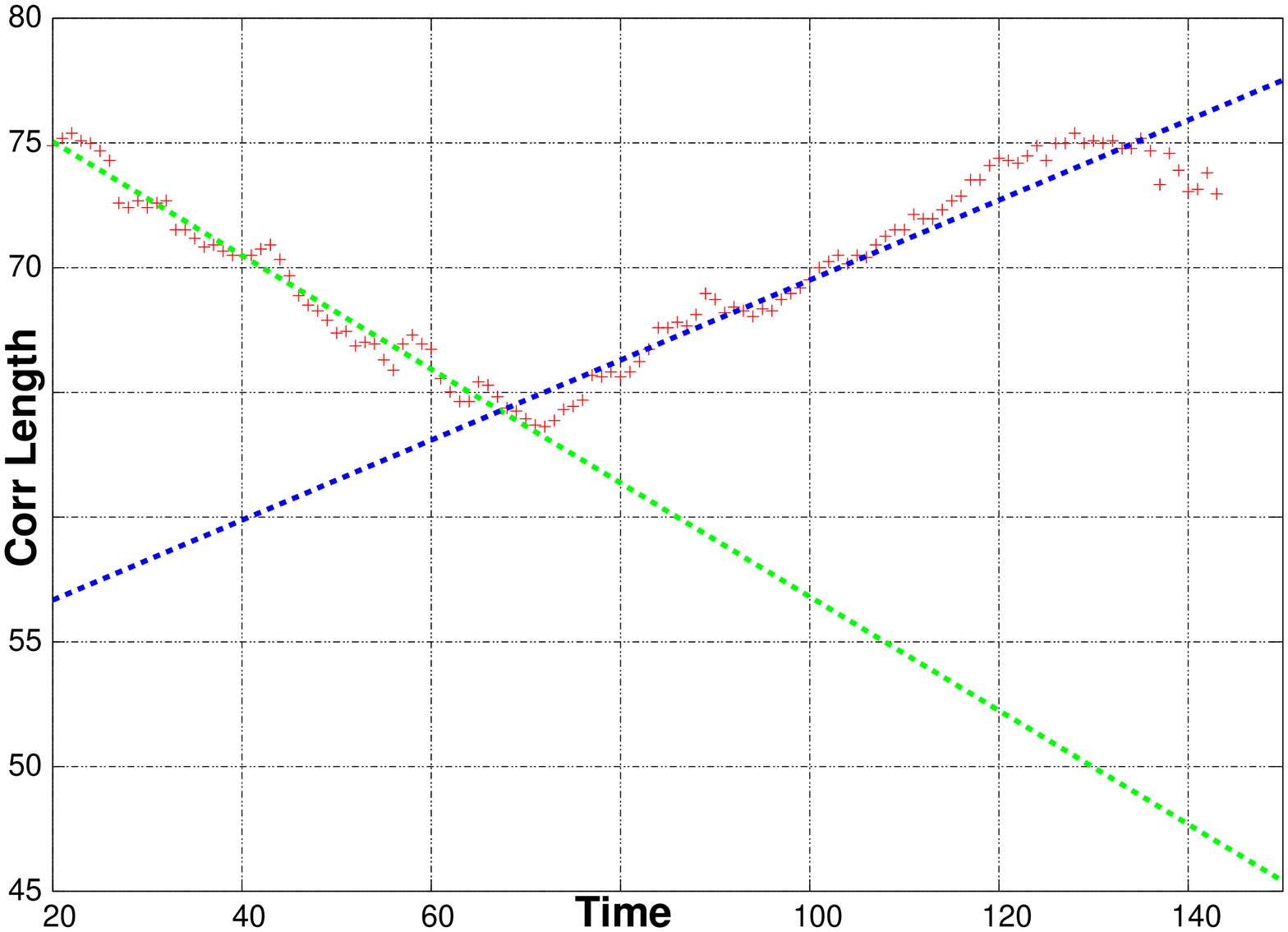}
    \caption{The Higgs (left), axion (middle), and bound state (right)
      string correlation length as a function of time. The network is a
      local-local one with a small amount of bound states.  The data
      and linear fits for the two regimes are shown.
      \label{LL_Low}} 
  \end{center} 
\end{figure}

We then draw the corresponding plots for local-local networks, with a
large and a small amount of bound states in Fig.~\ref{LL_High} and
Fig.~\ref{LL_Low} , respectively. We see, from Fig.~\ref{LL_High}, that
for a local-local network with a large amount of bound states, all
three components scale (as for the local-global case), whereas the
change in correlation lengths during evolution is not really apparent
(as opposed with the local-global case).

In Fig.~\ref{LL_Low}, we plot the correlation length of the Higgs,
the axion and their bound states for a local-local network with a small
amount of bound states. Here we explicitly see that there is a
violation of the {\sl scaling} behaviour for the bound states, as a
result of the increase of their length with time. As we have already
mentioned earlier, this violation does not affect the entire network,
which indeed scales.

\begin{figure}[htbp] 
  \begin{center} 
    \includegraphics[width=0.32\textwidth,angle=0]{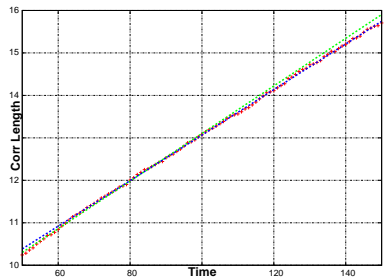}
    \caption{The correlation length for a single species of cosmic
      string.  There is no change in the slope of the correlation
      length as a function of time.
      \label{AH}} 
  \end{center} 
\end{figure}

In Fig.~\ref{AH} we show, for comparison, the correlation length of a
single species of string, using the same initial data as the one used
for the (first) Higgs of the $\pq$ strings. Clearly, there is no
change in the slope of the correlation length as a function of time.
This is an important check to assure that finite size effects do 
not influence the results of the simulations. 

\section{Conclusions}
In the context of brane-world models, brane collisions lead to the
formation of large Fundamental (F) strings and/or D1-branes
(D-strings). Those who survive the cosmological evolution become
cosmic superstrings, playing the r\^ole of their solitonic
analogues. By observing strings in the sky, we may be able to test for
the first (and maybe only) time string theory. To know the
observational consequences of cosmic superstrings, it is essential to
master the properties and evolution of such networks.

Performing numerical experiments, we have studied the dynamics and
overall properties of superstring networks.  More precisely, we have
performed field theory simulations of $p$ F- and $q$ D-strings, and
their $\pq$ bound states. We have investigated the effect of bound
states and in particular the approach to {\sl scaling}, which is
crucial for the cosmological consequences of cosmic superstring
networks. Defects which do not scale are cosmologically undesired,
since they may over-close the universe, leading to the {\sl old}
monopole problem.

Our studies have clearly shown that the three components of the network
scale, independently of the chosen initial conditions. In addition, by
having control over the initial abundance of bound states, we were
able to identify the effect of bound states on the overall network
evolution.  Finally, we have found that there is an additional energy
loss mechanism, beyond the chopping off loops. This new mechanism
consists on the formation of bound states, resulting to a lower  overall energy of
the network, thus leading to scaling.

\ack  The work of M.S. is partially supported
by the European Union through the Marie Curie Research and Training
Network {\sl Universenet} (MRTN-CT-2006-035863).

\end{document}